# Comment on "Nonlinearity of the Field Induced by a Rotating Superconducting Shell"

M. Tajmar

Space Propulsion & Advanced Concepts, Austrian Research Centers GmbH – ARC

A-2444 Seibersdorf, Austria

Abstract

An extension to the model from Berger is proposed to analyse the nonlinearity of the magnetic field from a rotating superconducting shell, taking into account the effect of the London penetration depth.

In a recent paper[1], Berger showed that the magnetic field produced from a rotating superconductor is not a strictly linear function of the angular velocity. Rather there exists a maximum field strength for the so-called London moment that can be calculated for a given radius and coherence length. Berger's model assumed a long superconducting cylindrical shell, where the thickness d is much smaller than the radius R and even small compared to the coherence length $\xi$. However, the model does not properly takes the effect of the London penetration depth $\lambda$ into account as it was shown by various papers[2-6], that the ratio of the shell thickness d with respect to the London penetration depth $\lambda$ has a significant influence on the London moment, especially for very thin film superconductors.

The Berger model used Ampere's law and the usual quantisation condition of the canonical momentum to derive the London moment. The basic equations are given as

$$B_i = B_o + \frac{4\pi}{c} N e^* v' \quad , \tag{1}$$

$$2\pi R m^* v + \frac{e^*}{c} \pi R^2 B_i = Lh \quad , \tag{2}$$

where $B_i$ and $B_o$ are the inner and outer magnetic field, N is the number of superconducting pairs per unit area, v' is their velocity relative to the ions of the shell, v is the velocity of the pairs relative to the laboratory and L is an integer that determines the trapped flux. Next, he assumed that the velocity of the Cooper-pairs in Eq. (2) is given by

$$v = v' + \omega R \quad , \tag{3}$$

where $\omega$ is the angular velocity. However, this can be considered only valid if the superconductor is thick with respect to the London penetration depth. As shown already in the first assessments by Becker[7] and London[8], the Cooper-pairs in the bulk will rigidly follow the lattice while the ones within the penetration depth will lag behind the lattice. This lag-current is then responsible for the magnetic field of the spinning superconductor. These two Cooper-pair regimes are correctly described by Eq. (3). However, what happens in the case of a superconductor that is thin with respect to the penetration depth? In this case, the rigid Cooper-pair current part is strongly decreased and thus the London moment field strength is reduced. Based on the corrections derived in previous papers[2-6], Eq. (3) shall be modified to

$$v = v' + \omega R \cdot \alpha \quad , \tag{4}$$

$$\alpha \cong \left(1 + \frac{2\lambda^2}{Rd}\right)^{-1} \quad . \tag{5}$$

By solving Eqs. (1), (2) and (5), we can now derive the London moment and the velocity of the Cooper-pairs as

$$B_i - B_o = \frac{2m^* c \gamma (\omega_\Phi - \alpha \omega)}{e^* (1 + \gamma)} \quad , \tag{6}$$

$$v = \frac{R[\gamma \omega + \omega_\Phi + (1 - \alpha) \omega]}{(1 + \gamma)} \quad , \tag{7}$$

using the same abbreviations as in the Berger model with

$$\gamma = \frac{2\pi(e^*)^2 RN}{m^* c^2}; \quad \omega_\Phi = \frac{Lh}{2\pi R^2 m^*} - \frac{e^* B_o}{2m^* c} . \quad (8)$$

The density of pairs N is obtained by minimizing the free energy following the Ginzburg-Landau equations. Our modified solution in Eq. (6) is changing the free energy solution of Berger to

$$G = \left(\frac{m^* cR}{e^*}\right)^2 \frac{\gamma}{2} \left( \gamma \omega_\kappa^2 - \omega_\xi^2 + \frac{\gamma(\omega_\Phi - \alpha\omega)^2 + (\omega_\Phi + (1 - \alpha + \gamma)\omega)^2}{(1+\gamma)^2} + \frac{2m}{m^*} \omega^2 \right) , \quad (9)$$

using again the same constants

$$\omega_\xi = \frac{\hbar}{m^* R\xi}; \quad \omega_\kappa = \frac{\kappa\hbar}{m^* d^{1/2} R^{3/2}} . \quad (10)$$

Note that for the case of $\alpha=1$, Eqs. (4), (6), (7) and (9) transform to the original Berger model equations.

A comparison with the results obtained in Berger's paper using the same assumptions ($\omega_\kappa/\omega_\xi = 0.1$, $\omega_\Phi=0$, R=1 cm, $\xi=10^{-4}$ cm) is shown in Fig. 1. It should be not surprising that our generalized model gives a slope which is much smaller than originally predicted. Berger's assumptions imply a penetration depth of $\lambda=3.2\times10^{-3}$ cm and a superconductor thickness of $d=10^{-7}$ cm. This reduces the initial slope 2 to a factor of 0.0061 as predicted from Eq. (5) in line with previous assessments[2-4].

Berger's model is restored if the thickness is at least comparable to the penetration depth. An example close to Berger's initial assumptions is an aluminium shell that still has a large coherence length of $\xi=10^{-4}$ cm but a smaller penetration depth of $\lambda=1.6\times10^{-5}$ cm. Fig. 2 shows a comparison for a thickness $d=5\times10^{-6}$ cm, close to the penetration depth, where the difference between the Berger and the generalized model is clearly seen. The initial slope is reduced by the correction factor in Eq. (5).

The following decay is stronger than just applying the correction factor to Berger's model, it is only correctly described by the full free energy model in Eq. (9). Thus, the generalized model further adds to the nonlinearity of the London moment originally proposed by Berger and should be taken into account when comparing data from precision experiments. In our assessment, we did not take into account the effect of an increasing penetration depth for the case of thin-film superconductors[9]. However, an increased $\lambda$ would favour even more the generalised model as the correction factor in Eq. (5) gets more dominant.

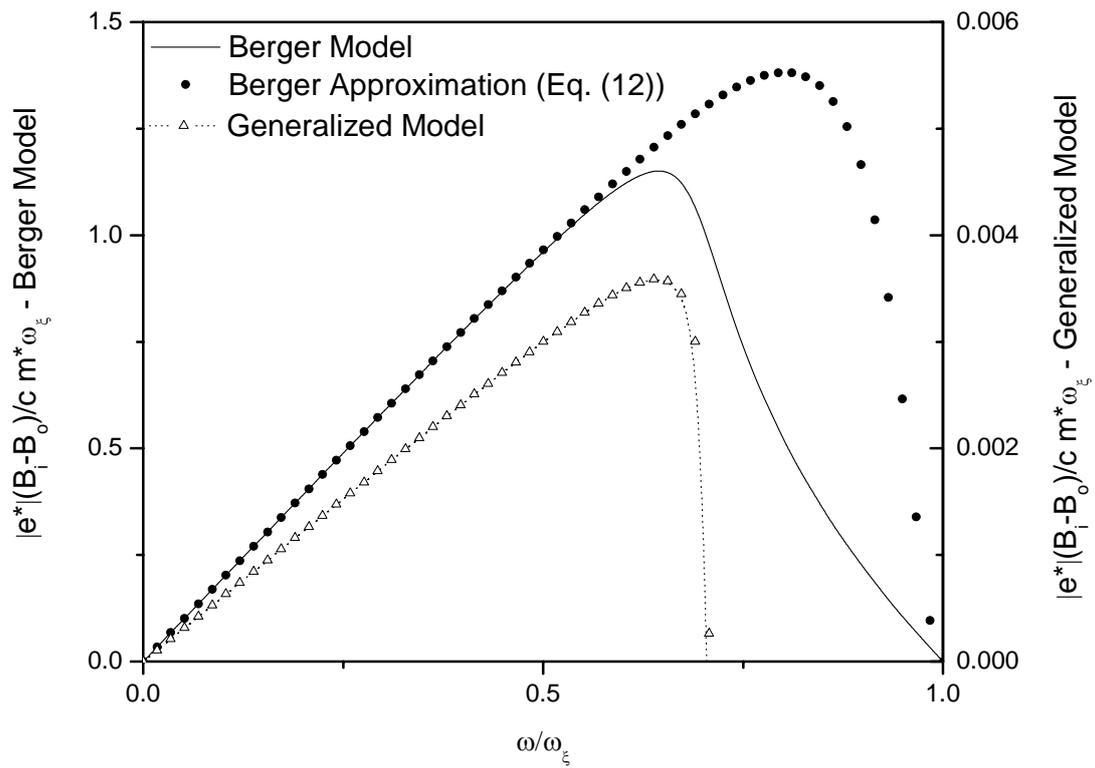

FIG. 1 Normalised magnetic field generated by the rotating shell versus angular velocity ($\omega_\kappa/\omega_\xi = 0.1$, $\omega_\Phi=0$, R=1 cm, $\xi=10^{-4}$ cm) similar to case analysed by Berger.

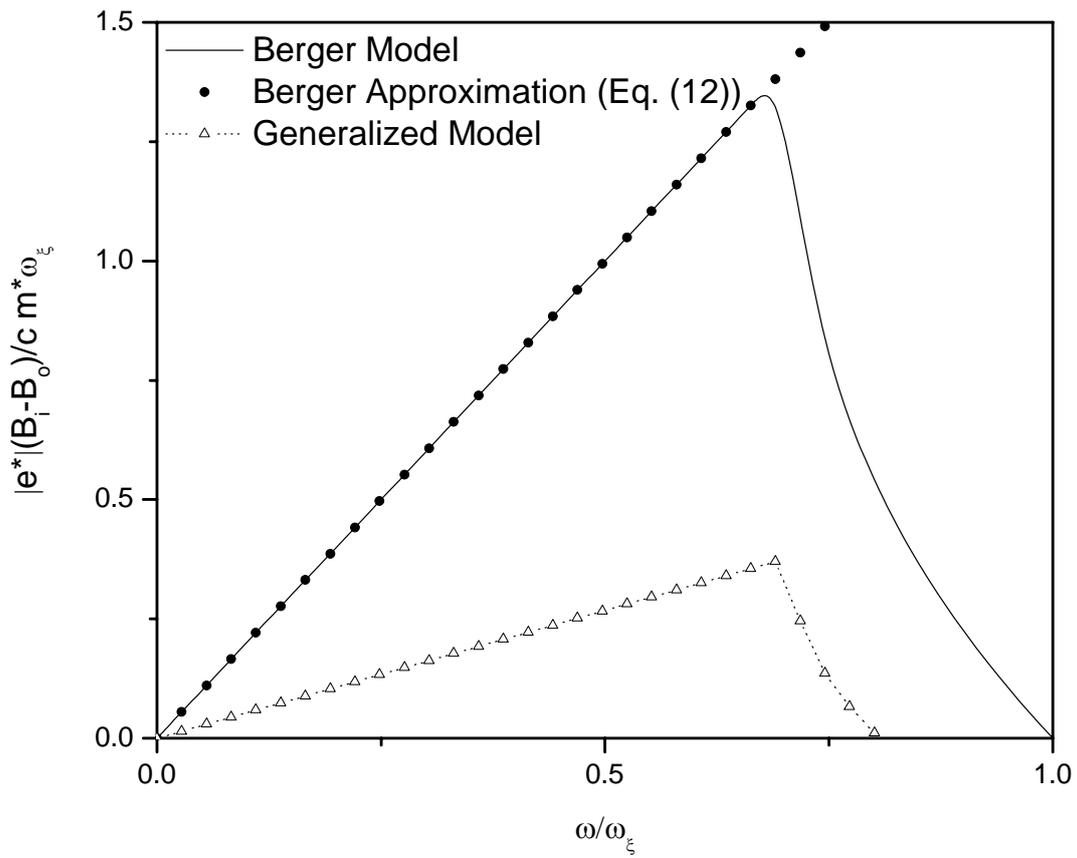

FIG. 2 Normalised magnetic field generated by the rotating shell versus angular velocity for Aluminum ($\omega_\kappa/\omega_\xi = 0.0022$, $\omega_\Phi=0$, R=1 cm, $\xi=1.6\times10^{-4}$ cm).